\documentclass[fleqn,usenatbib]{mnras}

\usepackage{newtxtext,newtxmath}
\usepackage{comment}

\usepackage[T1]{fontenc}

\DeclareRobustCommand{\VAN}[3]{#2}
\let\VANthebibliography\thebibliography
\def\thebibliography{\DeclareRobustCommand{\VAN}[3]{##3}\VANthebibliography}

\usepackage{graphicx}	
\usepackage{amsmath}	
\usepackage{physics}
\usepackage{float}
\usepackage{lipsum}
\usepackage{multirow}
\usepackage{xcolor}
\usepackage{orcidlink}
\usepackage{adjustbox}
\usepackage{subcaption} 
\usepackage{siunitx}
\usepackage{booktabs}
\usepackage{refcount}

\usepackage{xspace}
\newcommand\fluxratio{\ensuremath{\text{flux}_{\text{ratio}}}\xspace}
\newcommand\EWratio{\ensuremath{\text{EW}_{\text{ratio}}}\xspace}
\newcommand{\Oii}{\ensuremath{\left[\ion{O}{II}\right] \lambda3727}\xspace}
\newcommand{\Hb}{\ensuremath{\mathrm{H}\beta\,\lambda4861}\xspace}
\newcommand{\Oiiia}{\ensuremath{\left[\ion{O}{III}\right] \lambda4959}\xspace}
\newcommand{\Oiiib}{\ensuremath{\left[\ion{O}{III}\right] \lambda5007}\xspace}
\newcommand{\Ha}{\ensuremath{\mathrm{H}\alpha\,\lambda6563}\xspace}
 
\newcommand{\Niia}{\ensuremath{\left[\ion{N}{ii}\right] \lambda 6548}\xspace}
\newcommand{\Niib}{\ensuremath{\left[\ion{N}{ii}\right] \lambda 6583}\xspace}

\title[JWST NIRISS-NIRSpec Performance Analysis]{Quantifying Spectroscopic Flux Variations Between JWST NIRISS and NIRSpec: Slit Losses in Emission Line Measurements of z$\sim$1-3 Galaxies}

\author[Nicolò Dalmasso et al.]
{Nicolò Dalmasso$^{1,2}$\thanks{e-mail: ndalmasso@student.unimelb.edu.au}\orcidlink{0000-0002-1850-4050},
Peter J. Watson$^{3}$
\orcidlink{0000-0003-3108-0624},
Tommaso Treu$^{4}$
\orcidlink{0000-0002-8460-0390},
Michele Trenti$^{1,2}$
\orcidlink{0000-0001-9391-305},
Benedetta Vulcani$^{3}$
\orcidlink{0000-0003-0980-1499},
\newauthor Themiya Nanayakkara$^{5}$
\orcidlink{0000-0003-2804-0648},
Maru\v{s}a Brada\v{c}$^{6}$
\orcidlink{0000-0001-5984-0395},
Tucker Jones$^{7}$
\orcidlink{0000-0001-5860-3419},
Kristan Boyett$^{1,2,11}$
\orcidlink{0000-0003-4109-304X},
Xin Wang$^{8,9,10}$
\orcidlink{0000-0002-9373-3865},
\newauthor Sara Mascia$^{12,13}$
\orcidlink{0000-0002-9572-7813},
Laura Pentericci$^{12}$
\orcidlink{0000-0001-8940-6768}
\\
\\
$^{1}$School of Physics, University of Melbourne, Parkville, Vic 3010, Australia\\
$^{2}$ARC Centre of Excellence for All Sky Astrophysics in 3 Dimensions (ASTRO 3D), Australia\\
$^{3}$INAF -- Osservatorio Astronomico di Padova, Vicolo Osservatorio 5, 35122 Padova, Italy\\
$^{4}$Department of Physics and Astronomy, University of California, Los Angeles, 430 Portola Plaza, Los Angeles, CA 90095, USA\\
$^{5}$Centre for Astrophysics and Supercomputing, Swinburne University of Technology, PO Box 218, Hawthorn, VIC 3122, Australia\\
$^{6}$University of Ljubljana, Department of Mathematics and Physics, Jadranska ulica 19, SI-1000 Ljubljana, Slovenia\\
$^{7}$Department of Physics and Astronomy, University of California, Davis, 1 Shields Ave, Davis, CA 95616, USA\\
$^{8}$School of Astronomy and Space Science, University of Chinese Academy of Sciences (UCAS), Beijing 100049, China\\
$^{9}$National Astronomical Observatories, Chinese Academy of Sciences, Beijing 100101, China\\
$^{10}$Institute for Frontiers in Astronomy and Astrophysics, Beijing Normal University,  Beijing 102206, China\\
$^{11}$Department of Physics, University of Oxford, Denys Wilkinson Building,
Keble Road, Oxford OX1 3RH, UK\\
$^{12}$INAF – Osservatorio Astronomico di Roma, via Frascati 33, 00078 Monteporzio Catone, Italy\\
$^{13}$Institute of Science and Technology Austria (ISTA), Am Campus 1,
A-3400 Klosterneuburg, Austria\\
}

\date{Accepted XXX. Received YYY; in original form ZZZ}

\pubyear{2024}

\begin{document}
\label{firstpage}
\pagerange{\pageref{firstpage}--\pageref{lastpage}}
\maketitle

\begin{abstract}
We analyze JWST NIRISS and NIRSpec spectroscopic observations in the Abell 2744 galaxy cluster field. From approximately 120 candidates, we identify 12 objects with at least a prominent emission lines among \Oii, \Hb, \Oiiia, \Oiiib, and \Ha that are spectroscopically confirmed by both instruments. Our key findings reveal systematic differences between the two spectrographs based on source morphology and shutter aperture placement. Compact objects show comparable or higher integrated flux in NIRSpec relative to NIRISS (within 1$\sigma$ uncertainties), while extended sources consistently display higher flux in NIRISS measurements. This pattern reflects NIRSpec's optimal coverage for compact objects while potentially undersampling extended sources. Quantitative analysis demonstrates that NIRSpec recovers at least $63\%$ of NIRISS-measured flux when the slit covers $>15\%$ of the source or when $R_e<1$kpc. For lower coverage or larger effective radii, the recovered flux varies from $24\%$ to $63\%$. When studying the \Ha/\Oiiib emission line ratio, we observe that measurements from these different spectrographs can vary by up to $\sim$0.3 dex, with significant implications for metallicity and star formation rate characterizations for individual galaxies. These results highlight the importance of considering instrumental effects when combining multi-instrument spectroscopic data and demonstrate that source morphology critically influences flux recovery between slit-based and slitless spectroscopic modes in JWST observations.
\end{abstract}



\section{Introduction }\label{sec: introduction}

Spectroscopic observations are fundamental to astronomical research, enabling detailed investigations of celestial objects' physical properties, chemical compositions, and dynamical characteristics. Since the 19th century, when astronomers first used prisms and diffraction gratings to disperse starlight, advances in instrumentation have greatly expanded the range and sophistication of spectroscopic techniques available to researchers.

Modern astronomical instrumentation has developed three primary spectroscopic techniques that exemplify the sophisticated technological landscape of contemporary research: integral-field, slitless, and slit-based spectroscopy. Integral-field spectroscopy (IFS) provides the unique capability to obtain spatially resolved spectral data across extended objects. However, it presents significant technical challenges, including complex data reduction processes and substantial computational demands for processing the resulting three-dimensional data cubes (\citealt{Boker_2022}). The achievable spectral resolution with IFS is determined by the specific instrument design and can be comparable to that of traditional slit-based spectrographs, rather than being inherently limited by the technique itself.

Slitless spectroscopy, as implemented by the Wide Field Slitless Spectroscopy (WFSS) mode of the Near Infrared Imager and Slitless Spectrograph (NIRISS) aboard the James Webb Space Telescope (JWST, \citealt{Gardner_2006}), offers a comprehensive approach by dispersing light across the entire field of view, facilitating simultaneous spectral mapping of multiple objects. This method provides a broad observational perspective, capturing spectral information without the constraints of pre-selection, though it introduces the challenge of potential spectral overlap between sources and higher backround noise. The WFSS mode is particularly advantageous for multi-object observations and will serve as our primary focus when examining NIRISS capabilities in this work.

In contrast, slit-based spectroscopy, represented by the Near Infrared Spectrograph (NIRSpec, \citealt{Jakobsen_2022}), employs a more targeted observational strategy. While NIRSpec offers Fixed Slits Spectroscopy (FS), Multi-Object Spectroscopy (MOS) and Integral Field Unit (IFU) capabilities, this study will specifically examine the MOS mode employing a Micro-Shutter Assembly (MSA) narrow slit mask. This technique significantly reduces sky background contamination and enhances overall observational sensitivity. The precision of MSA-based spectroscopy allows for highly selective spectral observations, strategically isolating specific astronomical sources through carefully controlled shutter configurations. For an in-depth analysis of the MOS technique, we refer the reader to \cite{Ferruit_2022}.

A fundamental limitation of slit-based spectroscopy is the occurrence of slit losses. This effect arises when a fraction of the incoming radiation from a target source does not propagate through the entrance slit and is consequently excluded from the final recorded spectrum. Slit losses originate from a combination of geometric misalignment, the intrinsic source morphology, and the wavelength-dependent optical response of the instrument.

They are strongly dependent on the spatial extent and surface brightness distribution of the source, the point spread function (PSF) at each wavelength, and the on-sky alignment of the source relative to the slit. These losses introduce systematic uncertainties in the flux calibration, particularly for resolved or partially resolved sources, and must be accounted for either through empirical correction strategies or forward modeling of the instrumental response. Empirical corrections are generally based on available photometry or simple assumptions about the surface brightness distribution of the source. However, such corrections are not necessarily accurate for emission lines maps which are often not a simple scaling of the continuum surface brightness distribution and can be rather irregular.

In this study, we perform an initial exploration of the effects of slit losses using actual JWST NIRISS and NIRSpec data. In practice,
we compare flux measurement performance between JWST's NIRISS WFSS and NIRSpec MSA instruments operating in multi-object spectroscopy modes. This analysis examines how instrumental differences affect spectroscopic accuracy across varying observational parameters. We have explicitly limited our investigation to these multi-object observation modes, excluding NIRSpec's IFU capability as it falls beyond the scope of this work. The study quantifies performance disparities between these complementary spectroscopic approaches, with particular emphasis on slitloss effects in NIRSpec MSA relative to NIRISS WFSS observations. 

This work fulfills one of the key deliverables of The GLASS JWST Early Release Science Program (GLASS-JWST-ERS)\footnote{\url{https://archive.stsci.edu/hlsp/glass-jwst}\label{note: glass}} by providing a direct comparison between the two spectrographs. Our analysis is constrained by the unique characteristics of our dataset, which is based on the catalog release containing redshifts and emission line fluxes determined from JWST/NIRISS Wide-Field Slitless Spectroscopy observations in the Abell 2744 cluster field \citep{Watson_2025}. This catalog represents another important deliverable of the GLASS-JWST program and provides the foundation for our comparative spectroscopic study. We emphasize that, in the spirit of the Early Release Science Programs, this is just the first step in this investigation. A valuable future step will be to extend our investigations to larger samples of galaxies with both NIRSpec and NIRISS investigations. However, this next step will likely require some time, owing to the inherent complexity of the NIRISS dataset and data reduction, making this initial exploration more valuable.

The structure of this paper is outlined as follows: in Sec.\ref{sec: data selection}, we describe data reduction and the creation of the parent sample; in Sec.\ref{sec: Slit loss analysis}, we present two distinct analyses: precise emission line fitting and the study of emission line maps, with corresponding discussion of both approaches; and in Sec.\ref{sec: summary}, we offer a summary of our key findings. When needed we adopt the cosmological parameters determined by the Planck Collaboration \citep{Ade_Planck_2015}: $(\omega_{\rm{m}}, \Omega_{\lambda}, h, \sigma_{8}) = (0.3075, 0.6925, 0.6774, 0.8159)$. Magnitudes are reported in the AB system \citep{Oke_1983}.

\section{Data sets and sample selection}\label{sec: data selection}

\begin{figure*}
  \includegraphics[width=1\linewidth]{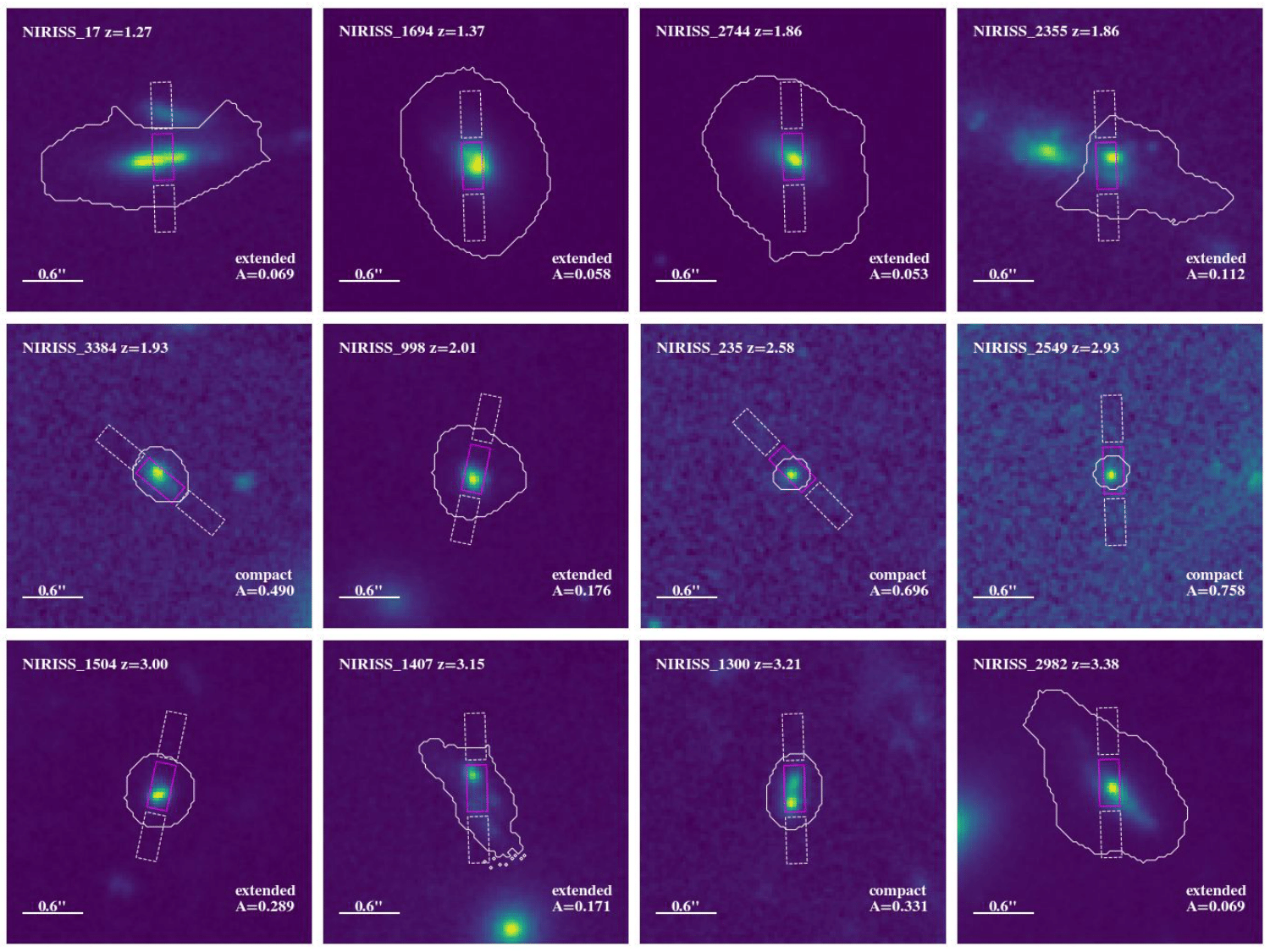}
\caption{NIRISS imaging cutouts are presented using a drizzled stack of the F115W, F150W, and F200W filters, processed to better highlight the light associated with each galaxy candidate. Solid white contours represent the NIRISS segmentation maps, rectangles show the NIRSpec slits assigned to the target (solid magenta and dash-dot white). The ID and redshift $z$ refer to measurements from NIRISS, using the catalog presented in \citet{Watson_2025} and a 0.6 arcsec scale bar is included. The "A" parameter, shown in the bottom right corner, represents the fractional area of the segmentation map that is encompassed by the slit. Objects are classified as either "compact" or "extended" based on the classification criteria presented in Sec.\ref{sec: data selection}.}
\label{fig: all cutouts}
\end{figure*}

\begin{table*}
\begin{adjustbox}{width=\textwidth,center}
\begin{tabular}{cccccccccccc}
    \midrule
    \multicolumn{12}{c}{Selected Candidates} \\
    \midrule
    \multicolumn{1}{c}{JWST Program} & \multicolumn{1}{c}{RA} & \multicolumn{1}{c}{DEC} & \multicolumn{1}{c}{NIRISS ID} & \multicolumn{1}{c}{NIRSpec ID} & \multicolumn{1}{c}{$z_{\rm{spec}}$} & \multicolumn{1}{c}{$m_{\rm{AB}}$} & \multicolumn{1}{c}{\Oii} & \multicolumn{1}{c}{\Hb} & \multicolumn{1}{c}{\Oiiia} & \multicolumn{1}{c}{\Oiiib} & \multicolumn{1}{c}{\Ha} \\
    (1) & (2) & (3) & (4) & (5) & (6) & (7) & (8) & (9) & (10) & (11) & (12)\\ 
    \midrule

    GLASS[1324] & 3.58625 & -30.41654 & 17 & 410067 & 1.27 & 23.3 & - & - & X & X & X \\
    GLASS[1324] & 3.57674 & -30.39360 & 1694 & 20021 & 1.37 & 22.71 & - & - & X & X & X \\
    GLASS[1324] & 3.57018 & -30.38372 & 2744 & 20008 & 1.86 & 22.81 & X & X & X & X & X \\
    GLASS[1324] & 3.58697 & -30.38699 & 2355 & 20025 & 1.86 & 24.41 & - & X & X & X & X \\
    UNCOVER[2561] & 3.59938 & -30.37594 & 3384 & 26882 & 1.93 & 26.35 & - & - & X & X & X \\
    DDT[2756] & 3.59852 & -30.40177 & 998 & 20015 & 2.01 & 24.79 & - & - & X & X & X \\
    UNCOVER[2561] & 3.58490 & -30.41102 & 235 & 8313 & 2.58 & 27.73 & - & - & X & X & - \\
    GLASS[1324] & 3.60686 & -30.38555 & 2549 & 90155 & 2.93 & 27.9 & - & X & X & X & - \\
    DDT[2756] & 3.60711 & -30.39562 & 1504 & 90012 & 3.00 & 25.01 & - & - & X & X & - \\
    GLASS[1324] & 3.57497 & -30.39678 & 1407 & 341568 & 3.15 & 24.2 & - & - & X & X & - \\    
    GLASS[1324] & 3.60780 & -30.39805 & 1300 & 341060 & 3.21 & 25.63 & - & - & X & X & - \\
    GLASS[1324] & 3.57786 & -30.38118 & 2982 & 360005 & 3.38 & 23.0 & X & - & - & - & - \\
    
    \midrule
\end{tabular}
\end{adjustbox}
\caption{This table summarizes the galaxies used in this study for slit loss characterization, as discussed in Sec.\ref{sec: data selection}. Columns (8)-(12) list the emission lines analyzed in this work, with an "X" indicating lines that are matched by both the NIRISS and NIRSpec instruments and for which we successfully fitted the emission line to calculate the integrated flux. Column descriptions: (1) JWST Program; (2) Galaxy RA [deg]; (3) Galaxy DEC [deg]; (4) Associated NIRISS ID (\citealt{Watson_2025}); (5) Associated NIRSpec ID (\citealt{Mascia_2024, Price_2024}); (6) Spectroscopic redshift from NIRISS; (7) Apparent AB magnitude from NIRSpec in the F200W filter.}
\label{tab:candidates}
\end{table*}

The NIRSpec spectral data used in this study were obtained through three public programs focused on the foreground galaxy cluster Abell 2744 and its immediate surroundings: (i) GLASS-JWST-ERS Program (PID 1324, PI T. Treu)\footnotemark[\getrefnumber{note: glass}], (ii) JWST-DDT Program (PID 2756, PI W. Chen), both with publicly available fully reduced spectra from \cite{Mascia_2024}, and (iii) UNCOVER-JWST Program (PID 2561, PI I. Labbé)\footnote{\url{https://jwst-uncover.github.io}\label{note: uncover}}, specifically the latest data release DR4, detailed in \cite{Price_2024}. 

The NIRSpec observations from GLASS utilized the high-resolution configuration with R2700 gratings, providing spectral resolutions of R$\sim$1400-3600 across the 0.7-5.0 $\mu$m range. In contrast, the DDT and UNCOVER programs employed the low-resolution R100 prism mode, delivering more moderate spectral resolutions of R$\sim$30-300 across the 0.6-5.3 $\mu$m wavelength range. We highlight an important distinction in NIRSpec data processing across our survey datasets. Data from GLASS-JWST-ERS and JWST-DDT programs retain path loss and barshadow corrections within their reduction pipeline, while UNCOVER-JWST DR4 data omits these corrections. The path loss correction addresses flux attenuation resulting from geometric aperture effects and diffraction losses, with particular significance for sources displaced from optimal slit positioning. The barshadow correction accounts for photon losses due to microshutter bar occlusion in MOS observations, providing necessary flux calibration for point source measurements after one-dimensional spectral extraction.

The NIRISS spectral data analyzed in this work were acquired through the publicly available GLASS-JWST-ERS Program, utilizing the catalog presented in \cite{Watson_2025}. These observations were obtained with the WFSS mode, using the F115W, F150W, and F200W blocking filters to cover a wavelength range $\approx1.0-2.2\,\mu$m at a spectral resolution of $R\approx150$ at 1.4 $\mu$m.
The wavelength coverage of these filters does not overlap, and the throughput is only greater than 50\% of the peak value within the windows 1013-1283 \AA, 1330-1671 \AA, and 1751-2226 \AA\ respectively.
In the F115W and F150W filters, approximately 1.5 hours were dedicated to direct imaging, with a further 3 hours in each of the two orthogonal grism orientations.
Total exposure times in the F200W filter were half the duration.
Galaxies were detected using a stack of the direct imaging in all three filters, drizzled to a pixel scale of $0.03$ arcsec/pixel.
This was performed without an additional diffuse background subtraction step, and with a threshold 1$\sigma$ above the existing background after filtering with a matched convolution kernel, in order to better capture the extended intracluster light present in this field.
The NIRISS candidate identifications referenced throughout our analysis correspond to those in \cite{Watson_2025}, and for further details regarding the data reduction process, we direct readers to that work.

The entire sample of 1D spectra used in this work was extracted from the 2D data using the optimal extraction technique described by \citet{Horne_1986}. Specifically, for NIRISS, the extraction profiles were obtained by summing the 2D spectra along the dispersion axis, as described in \citet{Watson_2025}. For NIRSpec, a gaussian profile was fitted to the signal along the dispersion axis to perform the extraction, following the procedures detailed in \citealt{Mascia_2024, Price_2024}.

For our analysis, we focused on strong emission lines widely used as diagnostics of galaxy properties: \Oii, \Hb, \Oiiia, \Oiiib, and \Ha\footnote{Note that at the spectral resolution of NIRISS, it is not possible to separate \Ha\ from either \Niia\ or \Niib. Throughout this paper, what we refer to as \Ha\ more precisely denotes the blended complex of all three lines.}. Given the wavelength coverage of NIRISS, we identified the redshift interval $z\in [0.5,4.9]$ as the range where any of these emission features would be observable.

Both the NIRISS and NIRSpec catalogs were filtered to select sources within this redshift range that exhibited at least two of the emission lines of interest, each detected at a signal-to-noise ratio $S/N>5$. To ensure the robustness of our final sample, we conducted a detailed visual inspection of the spectra from both instruments, confirming the presence and strength of the emission lines and verifying their suitability for further analysis. We also examined the drizzled NIRISS detection image, as described above, to assess the morphological properties of each candidate galaxy.\\
\noindent Galaxies were then classified as compact if $50\%$ of their flux distribution was contained within $80\%$ of half the NIRSpec slit height (with a full slit height of 0.46 arcsec); otherwise, they were considered extended.\\
\noindent The complete sample of 12 candidate galaxies selected for this study is shown in Fig.\ref{fig: all cutouts}, with a summary of their properties and detected emission lines provided in Tab.\ref{tab:candidates}.

\section{Spectroscopic Measurement Characterization}\label{sec: Slit loss analysis}

\subsection{Characterization of Emission Line Fluxes and Equivalent Widths}\label{subsec: precise fitting}

\begin{figure*}
  \includegraphics[width=1\linewidth]{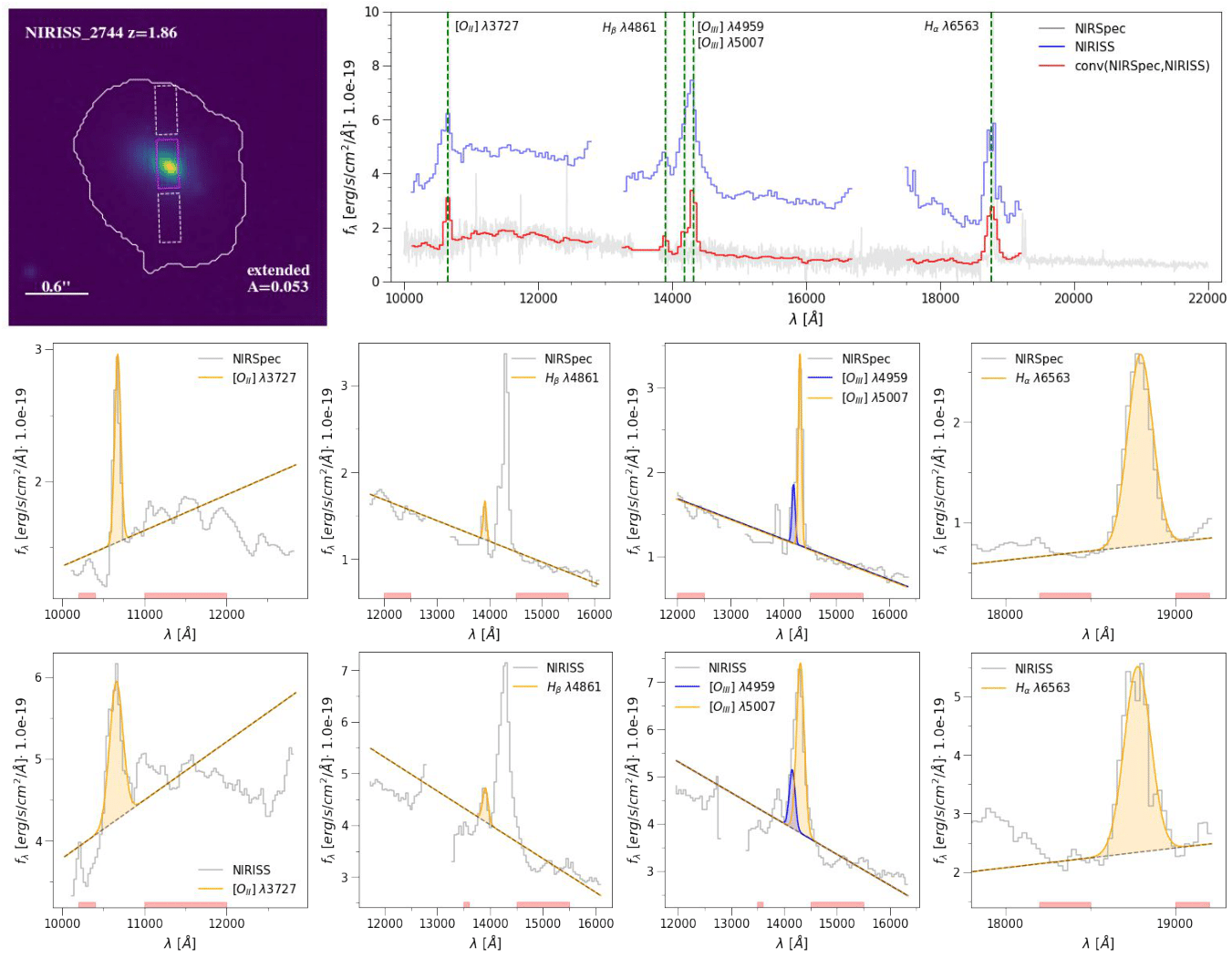}
\caption{Example Analysis: Spectroscopically Confirmed Candidate NIRISS ID 2744 at z=1.86. Top left: NIRISS image cutout (0.6 arcsec scale bar shown) with the NIRISS segmentation map (solid white contour) and NIRSpec slits assigned to the target (solid magenta and dash-dot white). Top right: Combined spectral data showing the original NIRSpec spectrum (gray), the NIRISS spectrum (blue), and the NIRSpec spectrum after convolution to match the NIRISS resolution (red), with identified emission lines marked. Bottom panels: Detailed emission line fits for all the line matched between the two instruments. The shaded red regions in each panel indicate an example of the continuum estimation regions computed for a single iteration of the MC sampling.}
\label{fig: spectra plots}
\end{figure*}

\begin{figure*}
  \includegraphics[width=1\linewidth]{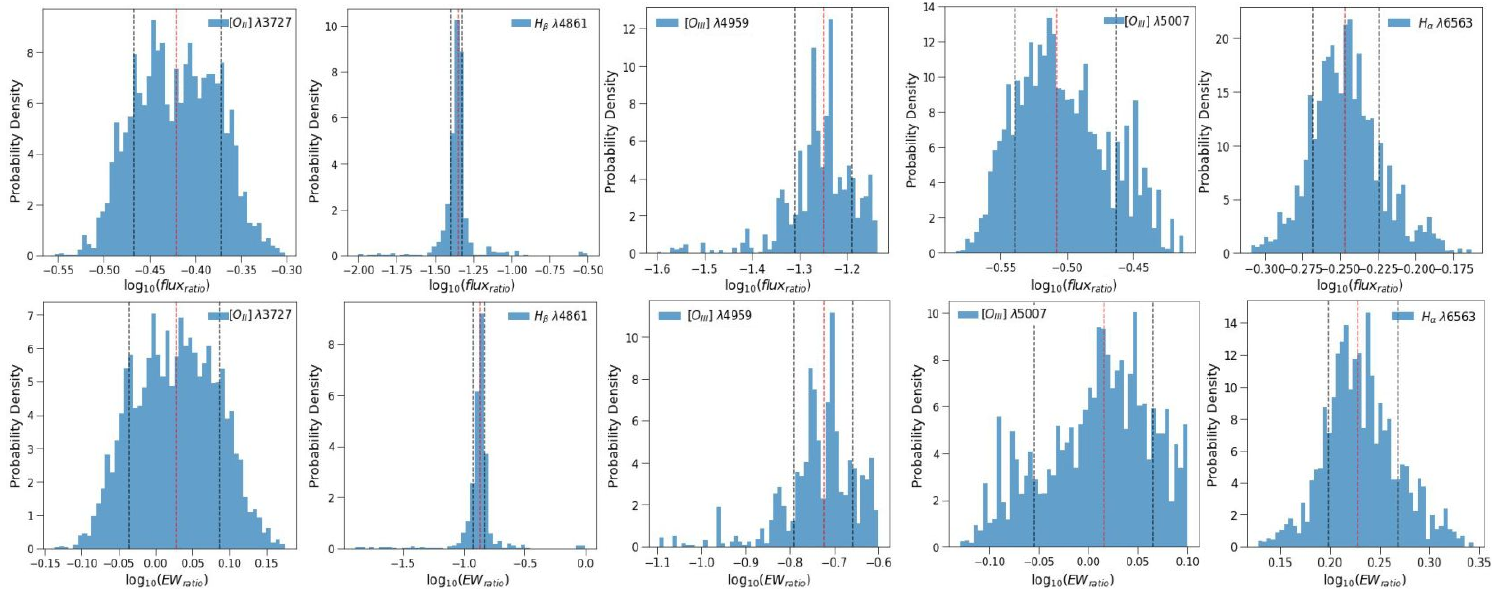}
\caption{Probability density distributions of \fluxratio (top row) and \EWratio (bottom row) for the emission lines of the source presented in Fig.\ref{fig: spectra plots}. The distributions are shown in logarithmic scale for visualization purposes and represent the marginalized probability densities obtained by varying the continuum estimation regions. In red the 50$^{\rm{th}}$ and from left to right in black 16$^{\rm{th}}$ and 84$^{\rm{th}}$ percentiles.}
\label{fig: candidate prob_dens}
\end{figure*}

For each galaxy in our sample, we conducted a systematic analysis of strong emission lines (detailed in columns (8)-(12) of Tab.\ref{tab:candidates}) using spectral data from both NIRISS and NIRSpec instruments.

The spectral analysis involved fitting emission lines through $\chi^2$ minimization to determine integrated fluxes from 1D spectra obtained from both NIRISS and NIRSpec instruments. We established flat continuum baselines by estimating levels from two regions surrounding each emission line of interest, incorporating Monte Carlo (MC) sampling of these continuum regions to account for uncertainties in continuum placement and obtain robust uncertainty estimates for the spectral line measurements.

The MC sampling consisted of $m^n$ total iterations, where $m=4$ denotes the number of continuum parameters varied (comprising the left and right mean distances from the emission line center, and the widths of both regions) and $n$ represents the number of MC iterations. Each emission line was inspected separately by assigning an initial region (both left and right of the emission line peak) from where to start the sampling. This was done to ensure that the initial region captured the characteristics of the spectrum in the vicinity of the line to be fitted, to avoid arbitrarily and systematically assigning exploration regions where there is a clear drop in the spectrum or a feature that would distort our analysis. Although the initial range differed for each emission line, a fixed minimum width of 100 Å was established for all regions, below which the MC sampling could not go. This resampling methodology enabled exploration of the parameter space for possible continuum definitions and assessment of how these variations influence our line measurements.

Fig.\ref{fig: spectra plots} shows one iteration from the $m^n$ MC sampling, illustrating the analysis performed for a representative galaxy. The top-left panel shows the galaxy's imaging, while the top-right displays the extracted spectrum, with measurements from NIRISS (blue) and NIRSpec (grey). To ensure consistent spectral resolution between instruments, we convolved the higher-resolution NIRSpec MSA data to match the lower-resolution NIRISS WFSS mode using a gaussian kernel with width corresponding to the quadrature difference between the resolution elements. The resulting convolved NIRSpec spectrum (red) was used throughout our comparative analysis to prevent resolution-dependent biases in line measurements and continuum features. The two subsequent rows illustrate a snapshot of the MC sampling of the continuum, highlighting the red-shaded regions on either side of the emission line used to define the continuum (in that precise snapshot), along with the corresponding emission line fit. The \ensuremath{\left[\ion{O}{III}\right]} doublet was fitted simultaneously, with the flux ratio \Oiiib/\Oiiia fixed at 3:1 according to quantum mechanical transition probabilities.

For each MC iteration and pair of matched emission lines, we measured the integrated flux and the equivalent width (EW) defining two ratios as:
\begin{subequations}\label{eq: all ratio}
\begin{align}
\text{flux}_{\text{ratio}} &= \frac{\text{flux}_{\text{NIRSpec}}}{\text{flux}_{\text{NIRISS}}} \label{eq: flux ratio} \\
\text{EW}_{\text{ratio}} &= \frac{\text{EW}_{\text{NIRSpec}}}{\text{EW}_{\text{NIRISS}}} \label{eq: EW ratio}
\end{align}
\end{subequations}

Fig.\ref{fig: candidate prob_dens} presents the resulting probability distributions of the integrated flux and EW ratios for each emission line when considering different continua. We adopted the 50$th$ percentile as our fiducial value, with 1$\sigma$ uncertainties derived from the 16th and 84th percentiles of these distributions.

\begin{figure*}
  \centering
  \begin{subfigure}[b]{0.49\textwidth}
    \includegraphics[width=\linewidth]{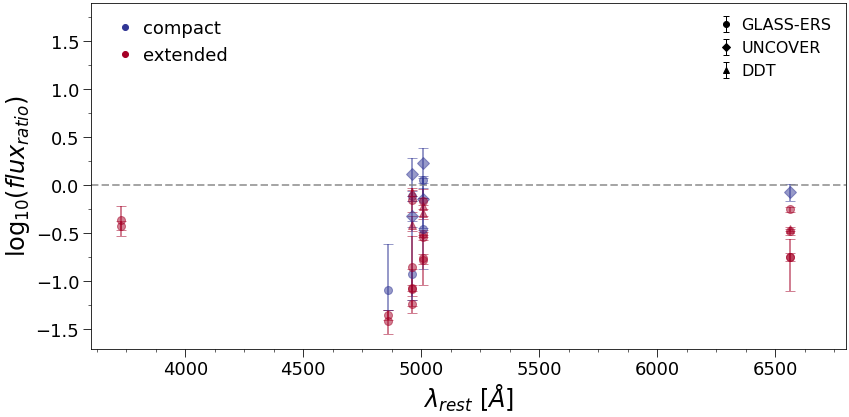}
    \caption{Logarithmic \fluxratio as a function of emission line rest wavelength.}
    \label{fig: flux_ratio vs rest_wave}
  \end{subfigure}
  \hfill
  \begin{subfigure}[b]{0.49\textwidth}
    \includegraphics[width=\linewidth]{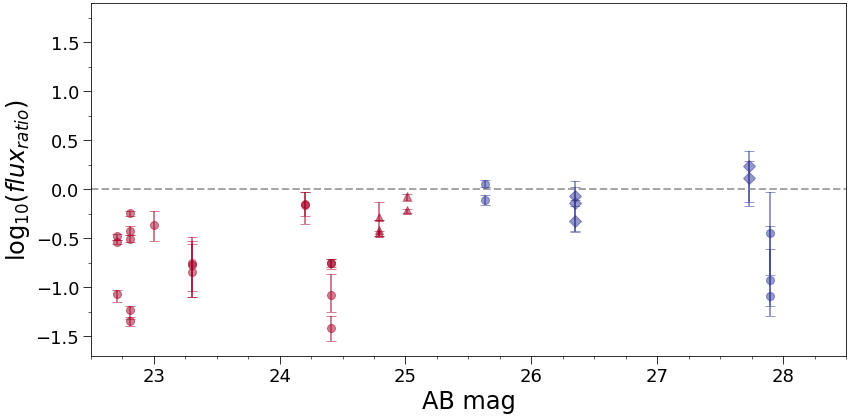}
    \caption{Logarithmic \fluxratio as a function of the galaxy AB magnitude.}
    \label{fig: flux_ratio vs mag}
  \end{subfigure}

  \vspace{0.5em} 

  \begin{subfigure}[b]{0.49\textwidth}
    \includegraphics[width=\linewidth]{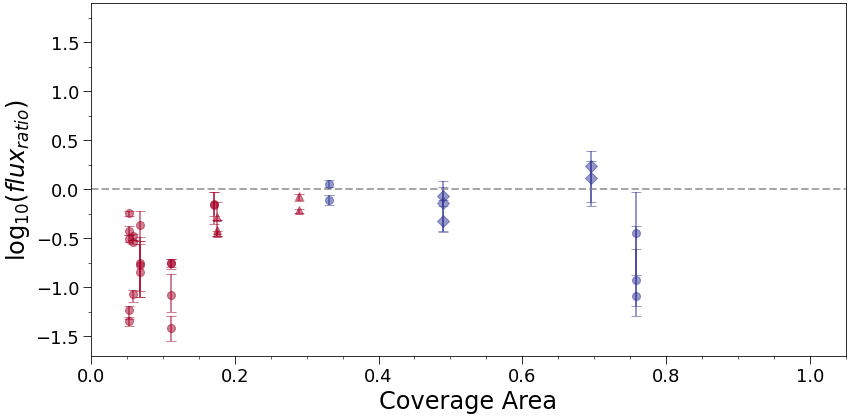}
    \caption{Logarithmic \fluxratio as a function of the fractional slit coverage area.}
    \label{fig: flux_ratio vs area}
  \end{subfigure}
  \hfill
  \begin{subfigure}[b]{0.49\textwidth}
    \includegraphics[width=\linewidth]{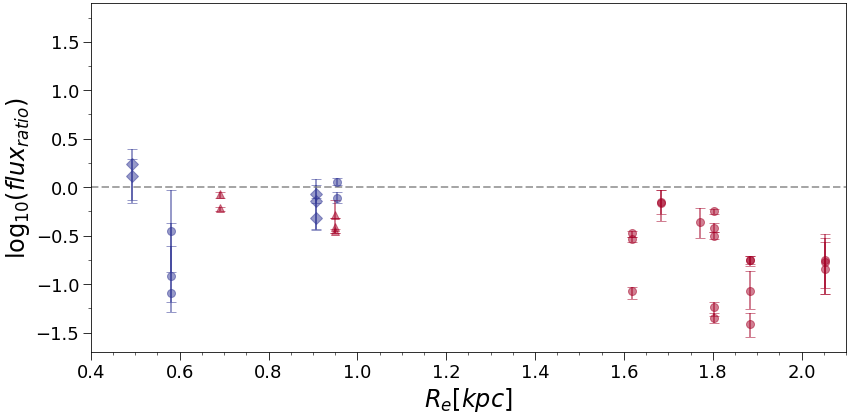}
    \caption{Logarithmic \fluxratio as a function of effective radius $R_e$ in $kpc$.}
    \label{fig: flux_ratio vs Re}
  \end{subfigure}

  \caption{
    All panels show results for the observed sources from three different JWST programs, color-coded by subsample (compact or extended). Error bars indicate 1$\sigma$ uncertainties calculated using the methodology described in Sec.\ref{subsec: precise fitting}. In these panels, we see that measurements for compact sources from NIRISS and NIRSpec are in good agreement. However, for more extended objects (shown on the left side of panel c), the NIRSpec measurements are systematically lower, as expected.
  }
  \label{fig: combined_flux_ratio}
\end{figure*}

\begin{figure*}
  \centering
  \begin{subfigure}[b]{0.49\textwidth}
    \includegraphics[width=\linewidth]{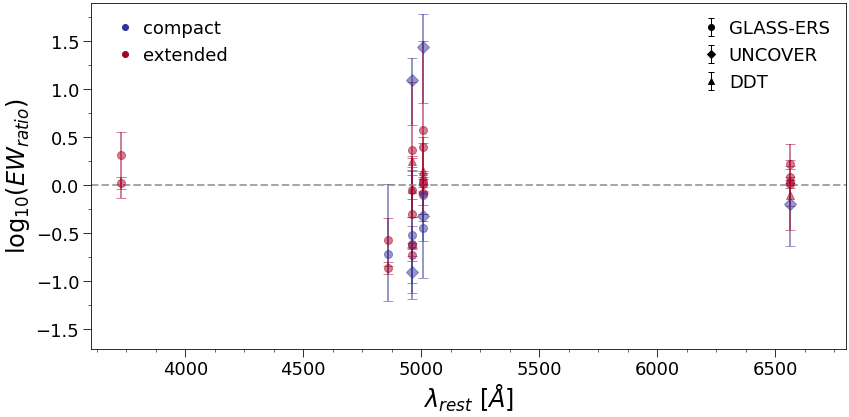}
    \caption{Logarithmic \EWratio as a function of emission line rest wavelength.}
    \label{fig: EW vs rest_wave}
  \end{subfigure}
  \hfill
  \begin{subfigure}[b]{0.49\textwidth}
    \includegraphics[width=\linewidth]{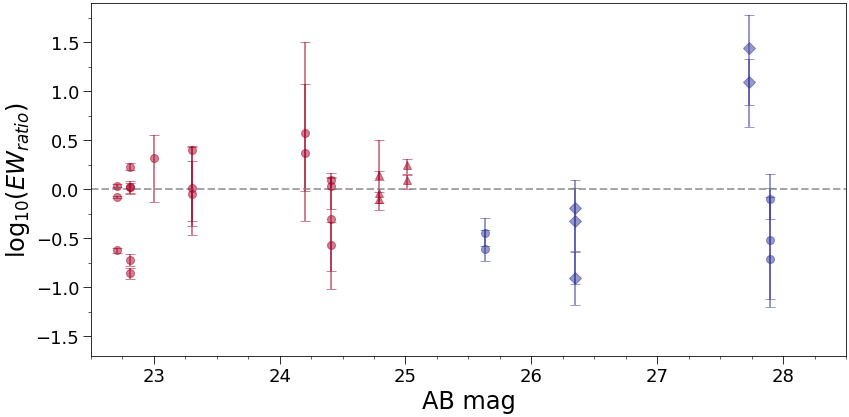}
    \caption{Logarithmic \EWratio as a function of the galaxy AB magnitude.}
    \label{fig: EW vs mag}
  \end{subfigure}

  \vspace{0.5em} 

  \begin{subfigure}[b]{0.49\textwidth}
    \includegraphics[width=\linewidth]{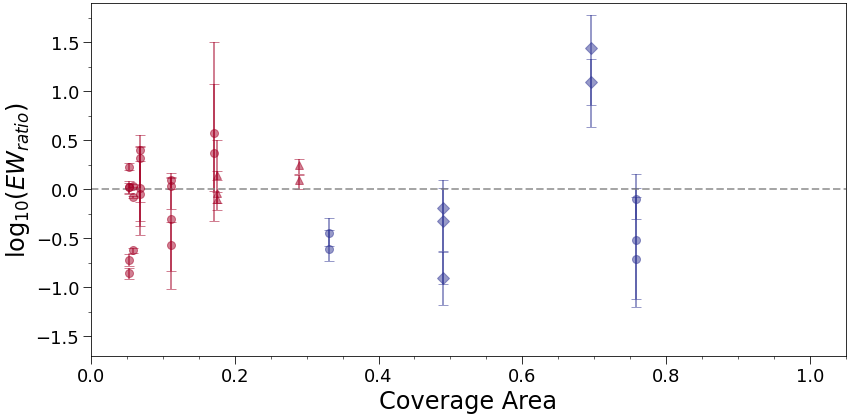}
    \caption{Logarithmic \EWratio as a function of the fractional slit coverage area.}
    \label{fig: EW vs area}
  \end{subfigure}
  \hfill
  \begin{subfigure}[b]{0.49\textwidth}
    \includegraphics[width=\linewidth]{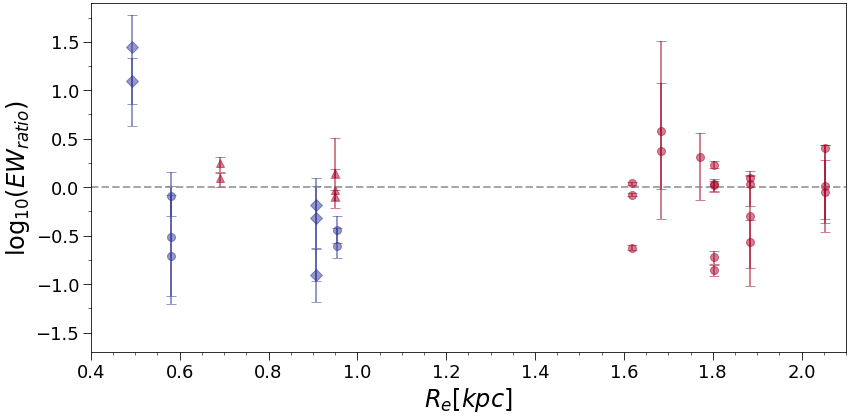}
    \caption{Logarithmic \EWratio as a function of effective radius $R_e$ in $kpc$.}
    \label{fig: EW vs Re}
  \end{subfigure}
  
  \caption{Same format as Fig.\ref{fig: combined_flux_ratio}, but showing the logarithmic \EWratio as a function of various quantities. Unlike the dichotomy seen in \fluxratio, both compact and extended sources display similar scatter in \EWratio with no strong trends.}
  \label{fig: combined_EW_ratio}
\end{figure*}

In both Fig.\ref{fig: combined_flux_ratio} and Fig.\ref{fig: combined_EW_ratio}, we plot against emission line rest-frame wavelength (panel a), galaxy AB magnitude in the F200W filter (panel b), NIRSpec slit coverage as a fraction of the NIRISS segmentation map area (panel c) and the effective radius of the galaxy $R_e$ (panel d). In all the panels we visually differentiate galaxies categorized as compact (blue) and extended (red) and with different markers indicating their observation program (see Sec.\ref{sec: data selection}).

Our analysis in Fig.\ref{fig: combined_flux_ratio} reveals a clear dichotomy between compact and extended sources. Compact sources show \fluxratio measurements with remarkable consistency, where NIRSpec integrated fluxes are comparable to or slightly exceed NIRISS measurements, agreeing within 1$\sigma$ uncertainties.

Extended sources exhibit systematically different behavior. NIRISS spectra consistently yield higher integrated flux measurements than NIRSpec observations, resulting in $\log_{10}(\fluxratio) < 0$. This persistent trend indicates a fundamental difference in how the two instruments measure flux for extended sources.

The systematic flux measurement differences observed between NIRISS and NIRSpec arise from their fundamentally distinct instrumental configurations, as shown in Fig.\ref{fig: all cutouts}. NIRSpec employs fixed-slit spectroscopy through a microshutter array (0.2 × 0.46 arcsec), confining observations to flux within narrow slits (magenta rectangles in Fig.\ref{fig: all cutouts}). While this configuration enables simultaneous multi-target observations with minimized background noise and maximized spectral resolution, it can systematically underestimate total flux for extended sources due to slit losses.

NIRISS, by contrast, utilizes slitless spectroscopy that disperses light across its entire field of view without spatial constraints. This approach captures the total emission within each galaxy's segmentation map (solid white contours in Fig.\ref{fig: all cutouts}), providing integrated flux measurements across the complete source morphology. The resulting systematic differences are particularly pronounced for extended galaxies, as demonstrated in Fig.\ref{fig: flux_ratio vs area} and Fig.\ref{fig: flux_ratio vs Re}, where larger sources show NIRISS flux measurements consistently exceeding those from NIRSpec.

These instrumental effects introduce systematic uncertainties that must be carefully considered when interpreting spectroscopic measurements. Recent studies have quantified these biases: \cite{He_2024} demonstrated that emission line fitting on 1D NIRISS spectra can systematically underestimate true line fluxes by approximately $30\%$ due to morphological broadening effects inherent to slitless spectroscopy, while \cite{Jiang_2024} quantified significant systematic uncertainties from NIRSpec slit losses through direct comparison with IFU data. The comparable integrated flux measurements between instruments for compact sources further validates this interpretation, as these targets are well-suited to NIRSpec's spatially constrained aperture.

On the other hand our analysis of the \EWratio presented in Fig.\ref{fig: combined_EW_ratio} reveals a different pattern compared to the \fluxratio analysis. While \fluxratio showed a clear dichotomy between compact and extended sources with persistent systematic offsets, the \EWratio exhibit more intricate  behavior.

For compact sources, EW ratios display substantial dispersion about around $\log_{10}(\EWratio)=0$, with several points deviating from it. This pattern diverges from the \fluxratio behavior observed in compact sources, where measurements exhibited closer adherence to the null hypothesis of equivalent flux recovery. The amplified scatter suggests that continuum-normalized measurements introduce systematic variability for compact objects that does not affect absolute flux determinations.

Extended sources exhibit a different behavior compared to their \fluxratio counterparts. While the logarithmic \fluxratio values were consistently negative (indicating that NIRISS systematically measured higher fluxes), the \EWratio measurements for extended sources are more scattered between approximately -1 and 0.5 in logarithmic scale. This difference indicates that continuum normalization mitigates the systematic flux differences observed between instruments, though morphological characteristics such as more or less clumpy regions within these extended sources likely contribute to the remaining scatter, as different instruments might sample these non-uniform structures differently. Several extended source measurements now appear consistent between instruments within uncertainties, which was not seen in \fluxratio analysis.

The wavelength panel (Fig.\ref{fig: EW vs rest_wave}) illustrates this distinction, where extended source \EWratio cluster closer to zero than their corresponding \fluxratio, especially when looking at longer restframe wavelengths. This suggests that while NIRSpec captures less total flux from extended sources due to slit limitations, the line to continuum ratio remains more comparable between instruments than absolute flux measurements would predict.

The \EWratio relationship with magnitude, coverage area and effective radius shows reduced systematic offsets compared to \fluxratio measurements, indicating that although NIRSpec captures less total flux from extended sources, the line-to-continuum ratio remains more consistent between instruments.

Our results suggest that EW measurements provide more consistent cross-instrument comparisons than direct flux measurements, particularly for extended sources. This improvement likely occurs because EW normalization by local continuum compensates for aperture-dependent light collection differences between NIRISS and NIRSpec. For spectroscopic studies using these instruments, flux measurements require calibration adjustments when analyzing extended sources, while EW-based analyses offer more reliable line strength comparisons, though wavelength-dependent effects should still be considered.

\subsection{Flux Characterization in Emission Line Maps}\label{subsec: emission line maps}

To further investigate the flux measurement differences between the two instruments, we focused on the impact of NIRSpec slit positioning on extended sources (defined in Sec.\ref{sec: data selection}, see also the cutouts in Fig.\ref{fig: all cutouts}).
For this analysis, we utilized the NIRISS emission line maps of \Oii, \Hb, \Oiiib, and \Ha. It should be noted that separate emission line maps for \Oiiia and \Oiiib were not available; instead, a single map represents the combined doublet emission \citep{Brammer_2023,Watson_2025}. 

We treated each galaxy under examination as a primary observation source, considering an optimal scenario for NIRSpec slit positioning. We used the galaxy coordinates (RA and DEC) from the catalog in Tab.\ref{tab:candidates} as reference. To account for potential uncertainties in aligning the slit with the galaxy's center, we sampled points from a gaussian distribution centered on these coordinates with $\sigma=0.12$ arcsec and further assumed precise NIRSpec slit positioning capability within a radius of $r=0.1$ arcsec around the slit's central point. This approach ensured plausible observations targeting bright galactic regions around the nominal center rather than peripheral points near the edge of the galaxy's segmentation map. For our MC sampling, we generated $n = 10^4$ simulated NIRSpec slits, allowing the slit rotation to vary randomly between $[0, \pi]$.

\begin{figure*}
  \includegraphics[width=1\linewidth]{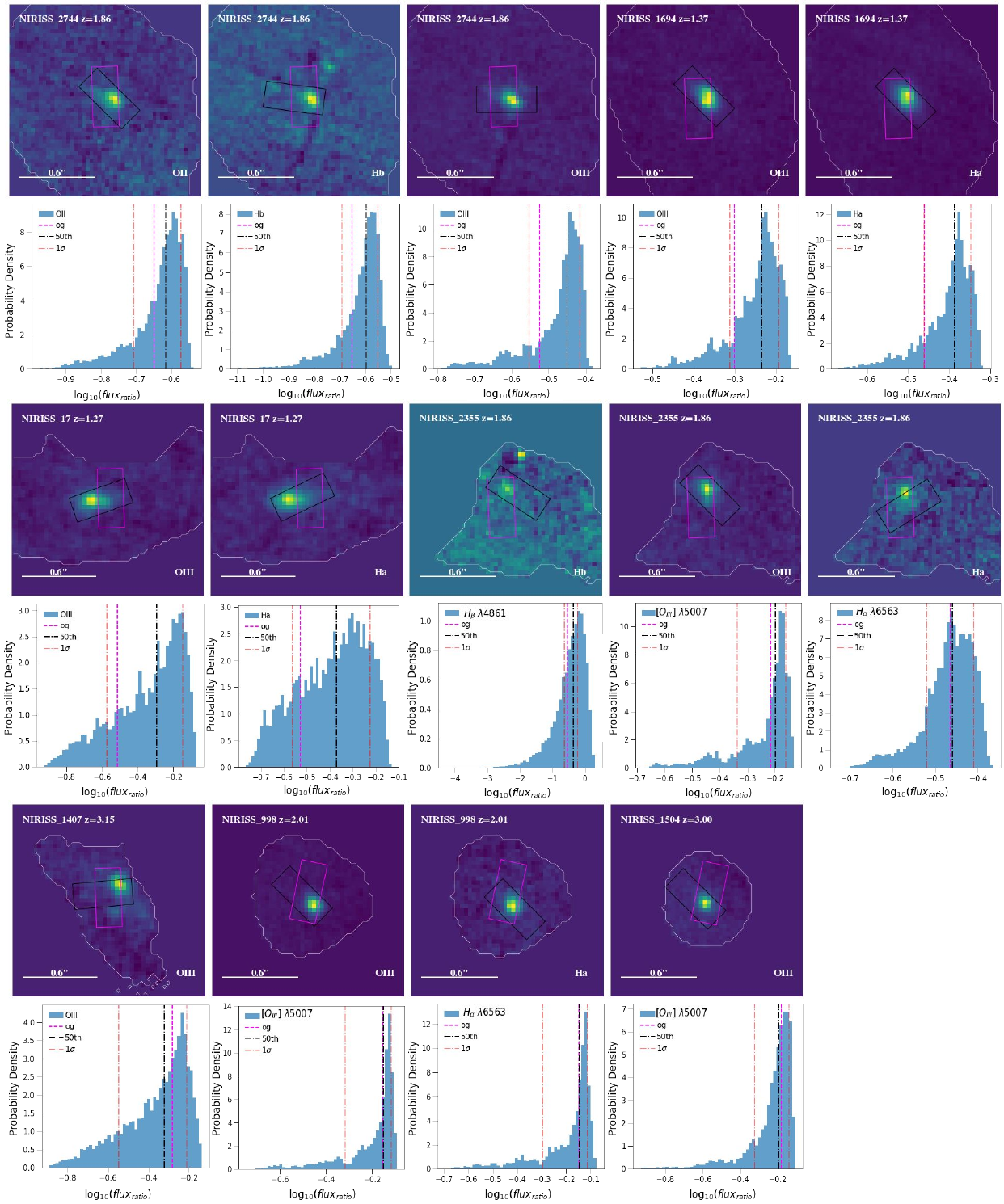}
\caption{NIRISS imaging cutouts from the corrisponding emission line maps for extended galaxies. Solid white contours represent NIRISS segmentation maps, while rectangles indicate NIRSpec slit positioning. Magenta (og: original) denotes the fixed position from the original observation from the corresponding JWST program. The black one represents the slit corresponding to the most probable \fluxratio (50th percentile) obtained from the MC analysis conducted on the associated emission line maps. For each cutout, the corresponding histogram displays the probability density of the \fluxratio (Eq.\ref{eq: flux ratio}) in logarithmic scale derived from the MC analysis and color-coded the same way with  associated $1\sigma$ uncertainties in red.}
\label{fig: emission_line_maps_plot}
\end{figure*}

Fig.\ref{fig: emission_line_maps_plot} presents the emission line map cutouts for each extended galaxy (solid white contour representing the NIRISS segmentation map), overlaid with the original NIRSpec slit (magenta) and the slit associated with the most probable (50$th$ percentile) \fluxratio measurements from the MC sampling (black). Each imaging panel is accompanied by the corresponding \fluxratio probability distribution obtained from the MC sampling, with color-coded vertical lines including the $1\sigma$ uncertainties. Numerical results from this analysis are presented in Tab.\ref{tab:flux_ratios_vs_area}.

The probability distribution panels clearly demonstrate that slit positioning is crucial for capturing the emitted flux from the observed galaxy. A more peripheral or uncentered positioning of the slit can significantly underestimate and compromise the spectral analysis, resulting in incorrect characterization of the source's redshift and brightness. This occurs because the slit captures only a fraction of the total flux, making measurements highly sensitive to the alignment between the slit and the galaxy's bright regions as observed in imaging data.

While the ideal approach would involve optimizing slit positioning and orientation for each individual galaxy, such a method is pragmatically unfeasible due to the substantial time constraints it would impose. Moreover, such an approach would fundamentally contradict the primary design philosophy of NIRSpec, which is engineered to enable simultaneous spectral observations of multiple astronomical targets.

Despite these limitations, our findings establish a quantitative framework for comparing flux measurements between NIRISS and NIRSpec. Given the assumptions required to approximate a realistic observation in the scenario when a slit is assigned to a candidate with an unfixed position and orientation in NIRSpec it is expected that flux measurements for an extended galaxy will likely be underestimated compared to those obtained with NIRISS. The extent of this underestimation is proportional to the size of the source relative to the MSA slit. 

\begin{figure*}
    \centering
    \begin{subfigure}[b]{0.49\textwidth}
        \includegraphics[width=\linewidth]{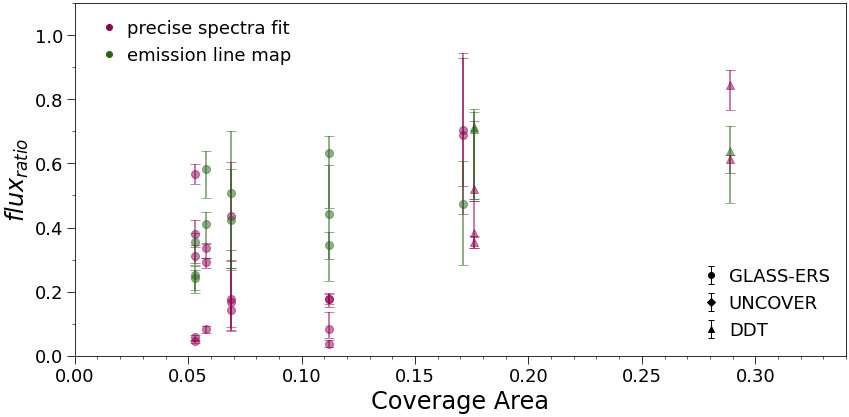}
        \caption{}
        \label{fig: flux_ratio vs area_extended}
    \end{subfigure}
    \hfill
    \begin{subfigure}[b]{0.49\textwidth}
        \includegraphics[width=\linewidth]{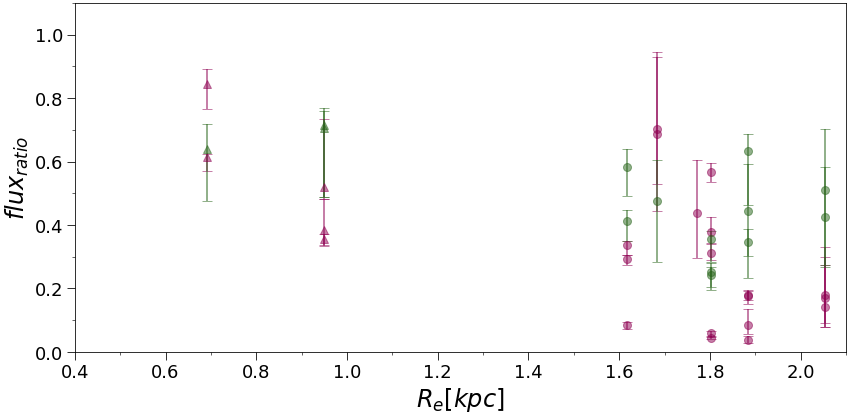}
        \caption{}
        \label{fig: flux_ratio vs Re_extended}
    \end{subfigure}
    \caption{\fluxratio as a function of slit coverage area and effective radius $R_e$ for extended galaxies.Results are color-coded by measurement method: "precise spectra fit" refers to measurements from the analysis in Sec.\ref{subsec: precise fitting}, while "emission line map" indicates results from the analysis in Sec.\ref{subsec: emission line maps} (also presented in Tab.\ref{tab:flux_ratios_vs_area}).}
    \label{fig: flux_ratio vs extended}
\end{figure*}

In the two panels of Fig.\ref{fig: flux_ratio vs extended}, we compare measurements of the \fluxratio for extended sources, showing results from the emission line maps (green) and the spectral fit (magenta) as a function of both coverage area and effective radius $R_e$. We find that when the galaxy area coverage exceeds $15\%$, or for sources with $R_e<1kpc$, the \fluxratio consistently remains above $63\%$, with both spectrographs yielding results that agree within their respective uncertainties. For coverage below $15\%$, corresponding to more extended sources, the recovered flux displays greater variation, ranging from $24\%$ to $63\%$. In this regime, emission line maps generally produce higher flux ratios compared to those derived from spectral fitting techniques.

We also examined the emission line ratio between \Ha and \Oiiib for four extended galaxies presenting these emission lines NIRISS[1694], NIRISS[2355] and NIRISS[998] from Fig.\ref{fig: emission_line_maps_plot}. We applied the previously detailed MC sampling method to investigate whether this line ratio maintains consistency across different galactic regions when observed using various methodologies, or if the different emission line maps exhibit intrinsic characteristics that cause the line ratio to vary rather than remain constant. 

\begin{figure*}
  \includegraphics[width=0.8\linewidth]{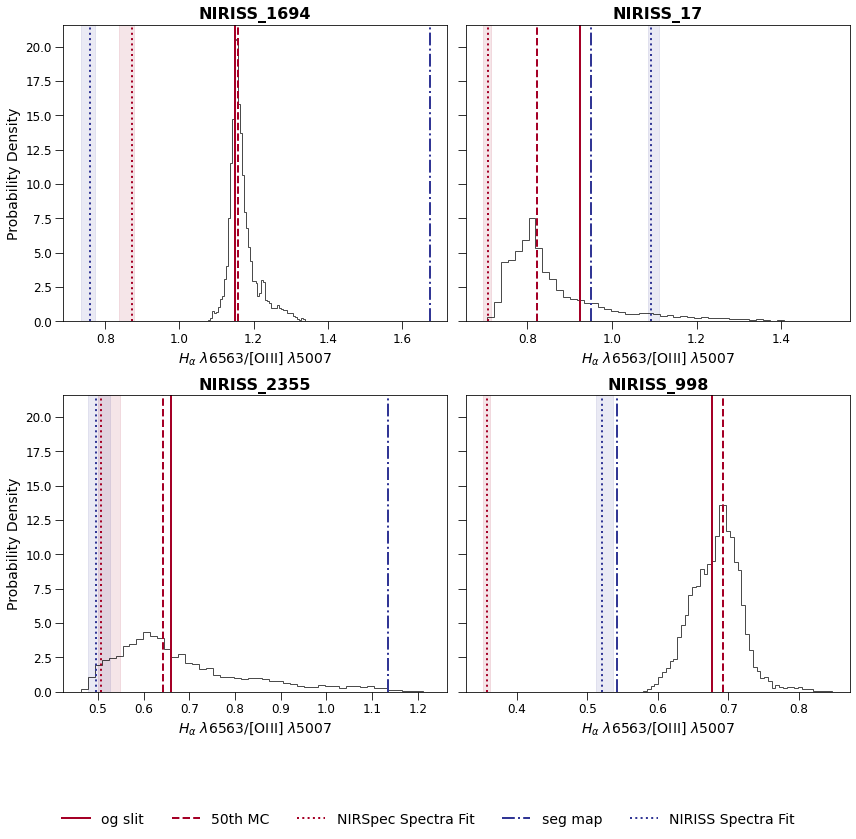}
\caption{Probability density histogram of the \Ha to \Oiiib emission line ratio for four extended sources. Multiple measurement methods are shown. NIRISS measurements: "seg map" represents the flux ratio measured within the segmentation map across both emission line maps. NIRSpec measurements: "og slit" shows flux measurements using the original assigned slit as the observed region, while "50th MC" indicates the 50$th$ percentile from MC sampling (Sec.\ref{subsec: emission line maps}). "Spectra Fit" for both instruments denotes the emission line ratio derived from precise emission line fitting detailed in Sec.\ref{subsec: precise fitting}.}
\label{fig: HaO3_emission_line_ratio}
\end{figure*}

Fig.\ref{fig: HaO3_emission_line_ratio} displays the probability distribution of the \Ha/\Oiiib line ratio across various measurement scenarios, including both emission line fitting using the 1D spectra (see Sec.\ref{subsec: precise fitting}) and measurements derived from emission line maps for each galaxy using both spectrographs. The vertical lines representing different instruments and methods for measuring the \Ha/\Oiiib emission line ratio are scattered across the probability distribution, producing inconsistent measurements that vary by $\sim0.3$ dex. Across all four panels, the measurements obtained from the precise spectral fit described in Sec.\ref{subsec: precise fitting} underestimate the probability distribution, providing lower values for the emission line ratio. While we expected the ratio to be similar within a margin of error regardless of measurement method or instrument, our findings suggest that line ratios are significantly influenced by observational position.

These observed inconsistencies in \Ha/\Oiiib line ratio measurements have critical implications for galactic parameter estimation. These variations could lead to substantial discrepancies in metallicity derivations depending on the instrument and methodology used (i.e., \citealt{Nagao_2006, Khoram_2025}), challenging the robustness of metallicity estimates, especially for dust-obscured systems where IR lines are often favored. For star formation analyses, the positive correlation between SFR and $H\alpha +H\beta$ luminosities means variations in line ratio measurements directly propagate into SFR uncertainties (\citealt{Kennicutt_1998}). This is further complicated by the finding that emission lines experience more attenuation than the continuum, with attenuation increasing with stellar mass, thus amplifying the impact of methodological differences on derived SFR and dust attenuation values (\citealt{Villa_2021}). 

Our results suggest that variations in line ratios are influenced by instrumental factors such as slit positioning, as well as data reduction and post-processing elements including segmentation map definition, contamination levels, and background attenuation. These methodological differences can artificially inflate uncertainties in both metallicity and SFR derivations. These findings underscore the need for standardized measurement protocols and careful cross-calibration when combining multi-instrument datasets, particularly in JWST studies comparing NIRISS and NIRSpec results.

\begin{table}
\centering
\resizebox{\columnwidth}{!}{
\begin{tabular}{cccccc}
  
    \midrule
    \multicolumn{6}{c}{\fluxratio from MC Sampling} \\
    \midrule
    \multicolumn{1}{c}{NIRISS ID} & \multicolumn{1}{c}{A} & \multicolumn{1}{c}{\Oii} & \multicolumn{1}{c}{\Hb} & \multicolumn{1}{c}{\Oiiib} & \multicolumn{1}{c}{\Ha} \\
    (1) & (2) & (3) & (4) & (5) & (6)\\ 
    \midrule
    2744 & 0.053 & 0.243$^{+0.023}_{-0.047}$ & 0.252$^{+0.027}_{-0.048}$ & 0.354$^{+0.028}_{-0.073}$ & - \\[.2cm]
    1694 & 0.058 & - & - & 0.584$^{+0.055}_{-0.097}$ & 0.411$^{+0.035}_{-0.063}$ \\[.2cm]
    17 & 0.069 & - & - & 0.509$^{+0.193}_{-0.241}$ & 0.425$^{+0.157}_{-0.152}$ \\[.2cm]
    2355 & 0.112 & - & 0.443$^{+0.15}_{-0.21}$ & 0.634$^{+0.053}_{-0.176}$ & 0.346$^{+0.044}_{-0.039}$ \\[.2cm]
    1407 & 0.171 & - & - & 0.475$^{+0.192}_{-0.131}$ & - \\[.2cm]
    998 & 0.176 & - & - & 0.708$^{+0.051}_{-0.213}$ & 0.712$^{+0.056}_{-0.235}$ \\[.2cm]
    1504 & 0.289 & - & - & 0.638$^{+0.079}_{-0.161}$ & - \\
    \midrule

\end{tabular}
}
\caption{Emission line \fluxratio in Eq.\ref{eq: flux ratio} obtained from different emission line maps of extended galaxies. Column descriptions: (1) Associated NIRISS ID; (2) Area of the galaxy covered by the slit; \fluxratio from emission line maps (3)-(6).}
\label{tab:flux_ratios_vs_area}
\end{table}

\section{Summary and Conclusions}\label{sec: summary}

This investigation examines 12 spectroscopically confirmed galaxies located in the Abell 2744 field, observed using both NIRISS and NIRSpec instruments on JWST. The dataset comprises observations from three JWST programs: GLASS-JWST-ERS (PID 1324), JWST-DDT (PID 2756), and UNCOVER-JWST (PID 2561). Our primary aim was to characterize spectroscopic measurement variations between these JWST spectrographs using both precise emission line fitting and Monte Carlo sampling-based analysis of emission line maps. Our key findings are:

\begin{itemize}

    \item Analysis of strong emission line fluxes revealed no significant trend in flux ratios with respect to rest-frame wavelength or AB magnitude of the observed sources.
    
    \item We identified a distinct dichotomy in \fluxratio between compact and extended sources. Compact objects showed consistent measurements between instruments, while extended galaxies exhibited systematically lower flux measurements with NIRSpec compared to NIRISS, attributable to the limited coverage of the NIRSpec MSA slit.
    
    \item EW ratios behaved differently than flux ratios. Compact sources showed greater scatter, but extended sources had better NIRISS-NIRSpec agreement for equivalent widths than for fluxes.
    
    \item Continuum normalization in EW calculations appears to mitigate the systematic offsets observed in direct flux comparisons, particularly for extended sources, suggesting that line-to-continuum ratios remain more consistent between instruments despite aperture differences.
    
    \item Our findings indicate that EW-based analyses may provide more reliable cross-instrument comparisons than direct flux measurements, especially for extended sources, though wavelength-dependent calibration effects should still be considered.
    
    \item Further investigation using emission line maps demonstrated that when the MSA slit covers at least 15\% of a galaxy's segmentation map area or the effective radius of the galaxy is $R_e<1kpc$, NIRSpec can recover a minimum of 63\% of the flux measured by NIRISS.
    
    \item Our analysis reveals that the \Ha/\Oiiib line ratio varies significantly ($\sim0.3$ dex) across measurement methods and instruments, potentially leading to discrepancies in metallicity and SFR estimates. This underscores the need for standardized protocols when interpreting spatially resolved galaxy properties from multi-instrument datasets.

\end{itemize}

\section*{Acknowledgements}
This research was supported in part by the Australian Research Council Centre of Excellence for All Sky Astrophysics in 3 Dimensions (ASTRO 3D), through project number CE170100013. This research was supported in part by The Dr Albert Shimmins Fund through the Albert Shimmins Postgraduate Writing Up Award (University of Melbourne). This work is based on observations made with the NASA/ESA/CSA James Webb Space Telescope. The data were obtained from the Mikulski Archive for Space Telescopes at the Space Telescope Science Institute, which is operated by the Association of Universities for Research in Astronomy, Inc., under NASA contract NAS 5-03127 for JWST. These observations are associated with programs JWST-ERS-1324, JWST-GO-2561, and JWST-DDT-2756. BV and PW acknowledge support from the INAF Large Grant 2022 “Extragalactic Surveys with JWST” (PI Pentericci), the INAF Mini Grant ``1.05.24.07.01 RSN1: Spatially-Resolved Near-IR Emission of Intermediate-Redshift Jellyfish Galaxies'' (PI Watson), and are supported  by the European Union – NextGenerationEU RFF M4C2 1.1 PRIN 2022 project 2022ZSL4BL INSIGHT. MB acknowledges support from the ERC Advanced Grant FIRSTLIGHT and Slovenian national research agency ARIS through grants N1-0238 and P1-0188.

\section*{Data Availability }

The datasets were derived from sources in the public domain:
\begin{itemize}
    \item GLASS-JWST ERS Program (PID 1324, PI T. Treu) and JWST
DDT Program (PID 2756, PI W. Chen) NIRSpec data in \cite{Mascia_2024} available at \url{https://archive.stsci.edu/hlsp/glass-jwst}

    \item GLASS-JWST ERS Program (PID 1324, PI T. Treu) NIRISS Spectroscopic Catalogue in \cite{Watson_2025} available at \url{https://archive.stsci.edu/hlsp/glass-jwst}
    
    \item UNCOVER-JWST Program (PID 2561, PI I. Labbé) Data Release 4 NIRSpec data in \cite{Price_2024} available at \url{https://jwst-uncover.github.io}
\end{itemize}


\bibliographystyle{mnras}
\bibliography{main} 


\appendix


\bsp	
\label{lastpage}
\end{document}